\documentclass[fleqn,usenatbib,useAMS]{mnras}

\usepackage{graphicx}	
\usepackage{amsmath}	
\usepackage{amssymb}	
\usepackage{multicol}        
\usepackage{bm}		
\usepackage{pdflscape}	

\usepackage[T1]{fontenc}
\usepackage{ae,aecompl}

\usepackage{newtxtext,newtxmath}
\DeclareUnicodeCharacter{2212}{-}

	\title[4U $1630-47$: Spectro-polarimetric analysis of an outburst ]{
	IXPE and NICER view of Black hole X-ray binary 4U $\textbf{1630-47}$:
    First significant detection of polarized emission in thermal state}
	\author[Ankur Kushwaha et al.]{
	Ankur Kushwaha$^{1,\,2}$\thanks{E-mail: ankurksh@ursc.gov.in},
	Kiran M. Jayasurya$^{1}$,
	Vivek K. Agrawal$^{1}$,
    Anuj Nandi$^{1}$
	\\
	$^{1}$Space Astronomy Group, ISITE Campus, U. R. Rao Satellite Centre,
	Outer Ring Road, Marathahalli, Bangalore, 560037, India\\
	$^{2}$Department of Physics, Indian Institute of Science, Bangalore, 560012, India\\
	}
	
	\date{Accepted XXX. Received YYY; in original form ZZZ}
	
	\pubyear{2023}
	
\begin{document}
	\label{firstpage}
	\pagerange{\pageref{firstpage}--\pageref{lastpage}}
    \maketitle
	
	\begin{abstract}
	We present a detailed spectro-polarimetric study of Black 
    hole X-ray binary 4U $1630-47$ during its $2022$ outburst 
    with {\it IXPE} and {\it NICER} observations. 
    The source is observed in disk dominated thermal state
    (kT$_{in}\approx1.4$ keV) with clear detection of absorption
    features at $6.69\pm0.01$ keV and $6.97\pm0.01$ keV from both 
    {\it NICER} as well as {\it IXPE} spectra, likely indicating
    a coupling of disk-wind. A significant degree of polarization 
    (PD) $= 8.33\pm0.17\%$ and polarization angle
    (PA) $= 17.78^{\circ}\pm0.60^{\circ}$
    in the energy range of $2-8$ keV are measured with 
    {\it IXPE}. PD is found to be an increasing function 
    of energy whereas PA remains roughly same within 
    the energy range. Simultaneous energy spectra from {\it NICER} 
    in the range of $0.5-12$ keV are modelled to 
    study the spectral properties. Furthermore, the spin parameter 
    of the black hole is estimated with spectro-polarimetric data 
    as a$_{\ast}=0.920\pm0.001\,(1\sigma)$ which is corroborated by {\it NICER} 
    observations. Finally, we discuss the implications of our findings. 
	\end{abstract}


	
	\section{Introduction}
	\label{sec: intro}

    The spectro-polarimetric study of the emission from 
    Black hole X-ray binaries (BH-XRBs) can reveal the geometry
    and dynamics of accretion processes. This emission can be of 
    thermal and non-thermal origins. 
	The Keplerian accretion disk \citep{1973A&A....24..337S} 
	is considered to produce multi-colour thermal X-ray emission 
	whereas inverse-comptonisation by a `hot' corona  
	\citep{1994ApJ...434..570T, 1995xrbi.nasa..126T, 1995ApJ...455..623C} 
	of soft photons emanating from disc is believed to be
    responsible for	higher energy non-thermal emission.
    These two types of emissions are the major components 
	in the energy spectra, of which one component may dominate 
    in a spectral state of BH-XRBs
    \citep[and references therein]{2001ApJS..132..377H,
	2005Ap&SS.300..107H, 2006ARA&A..44...49R,
	2012A&A...542A..56N, 2020MNRAS.497.1197B, 
    2021MNRAS.507.2602K}.

    The X-ray emission during soft state is expected to be 
    linearly polarized parallel to the disk plane as the 
    thermal emission suffers electron scattering in the 
    inner disc region. Similarly, the emission in hard state is 
    believed to be polarized as well but in the direction of normal 
    to the disk plane. Since the seed photons from the disk are 
    up-scattered within corona and produce polarized high energy 
    photons \citep{1977Natur.266..429S, 
    1980ApJ...235..224C, 2009ApJ...701.1175S}. 
    
    Apart from different spectral states, BH-XRBs are also known
    for their transient nature wherein the source exhibits
    sudden increase in flux after a prolonged quiescence phase 
    \citep{1995JBAA..105R.284L, 1997ApJS..113..367B}. 
    Such outbursts can last from weeks to months before the source 
    fades into quiescence. The frequency and duration of these 
    outbursts varies from source to source, or sometimes even 
    among different outbursts of the same 
    source itself \citep[and references therein]{2001ApJS..132..377H, 
    2005A&A...440..207B, 2012A&A...542A..56N, 
    2018JApA...39....5S, 2019MNRAS.487..928S}.
        
    4U $1630-47$ is a recurrent X-ray transient classified 
    as a black hole candidate from its spectral and timing properties
    \citep{1996ApJ...473..963B, 2005PASJ...57..629A}.
    It undergoes quasi-periodic outbursts with a recurrence time of 
    $\sim\!600$ days \citep{1995ApJ...452L.129P}.     
    It is located near Galactic plane and has a 
    high absorption column density with 
    N$_{H}\sim\!8\times10^{22}$ cm$^{−2}$ 
    \citep{2002ApJ...581..562S, 2010ApJ...713..257U}
    which makes it difficult to estimate the mass (M$_{\rm{BH}}$), 
    distance (D) and inclination ($i$) of the system. 
    The best estimation 
    of these parameters are M$_{\rm{BH}}\sim\!10$ M$_{\odot}$, 
    $D=10\pm0.1$ kpc and $i=60^{\circ}-70^{\circ}$ 
    \citep{1998ApJ...494..753K, 1998ApJ...494..747T, 
    2001A&A...375..447A, 2014ApJ...789...57S}. 
    {\it NuSTAR} observations suggest that 
    the source hosts a maximally rotating BH with 
    spin parameter: $a_{\ast}$ = $0.985^{+0.005}_{-0.014}$ 
    \citep{2014ApJ...784L...2K}. \citet{2018ApJ...867...86P} 
    obtained the same as $0.92^{+0.02}_{-0.01}$ using 
    {\it Chandra} and {\it AstroSat} observations of
    soft state during the $2016$ outburst. 
    
    4U $1630-47$, till date, has undergone more than $25$ 
    outbursts in the $\sim\!50$ years since its detection in $1969$. 
    At the time of writing this letter, the source is found 
    to be in another outburst. Moreover, polarization measurement 
    during an outburst of 4U $1630-47$, to the best of 
    our knowledge, has not been reported so far. 
        
    In this work, we focus on the spectro-polarimetric study 
    of 4U $1630-47$ during its outburst which started on 
    $28$ July, $2022$ and peaked on $21$ August, $2022$ 
    \citep{2022ATel15575....1J}. Subsequent to an alert generated by 
    {\it MAXI} \citep{2011PASJ...63S.623M}, 
    {\it NICER} \citep{2016SPIE.9905E..1HG} and 
    {\it IXPE} \citep{2022JATIS...8b6002W} carried out long     
    observations during the outburst under the 
    target of opportunity (ToO) campaign. 
    We make use of {\it IXPE} and {\it NICER} data 
    to study simultaneous polarimetric and spectral
    properties of 4U $1630-47$ in $2 - 8$ keV 
    and $0.5 - 12$ keV energy bands, respectively.  
 
	In $\S$\ref{sec: obs_data}, we provide details of the observations 
	and the steps for data reduction. The outburst profile along with
    spectro-polarimetric data modelling using both {\it NICER} as well as 
    {\it IXPE} are presented in $\S$\ref{sec: modelling and resuts}.
	Finally, we discuss the results and conclude in 
	$\S$\ref{sec: discussion}.    

	\section{Observations and Data Reduction}
 	\label{sec: obs_data}

    \begin{table}
    \centering
    \caption{Log of {\it NICER} and {\it IXPE} observations of 4U1630-47 
    considered in the analysis.
			From left to right,
            (1) name of observation;  
			(2) observation ID;
            (3) start MJD of observation;
            (4) exposure time
            (5) unabsorbed flux, in units 
		        of $\,\times 10^{-9}$ ergs cm$^{-2}$ s$^{-1}$
                in the mentioned energy range.}
	\label{tab: obs_table}
    \scalebox{0.85}{
    \begin{tabular}{ccccc}
    \hline
    \hline
    Obs Name & ObsID& Start MJD & Exposure (ksec)   & Flux\\
    \hline
    &\multicolumn{3}{c}{NICER}& $0.5-10$ keV \\
    \hline
    N1  & $5130010101$ & $59790$ & $\sim\!2.1$ & $5.67$ \\
    N2  & $5130010105$ & $59799$ & $\sim\!3.8$ & $6.75$ \\
    N3  & $5130010110$ & $59805$ & $\sim\!3.5$ & $8.93$ \\    
    N4  & $5501010102$ & $59814$ & $\sim\!2.0$ & $9.65$\\
    N5  & $5501010103$ & $59815$ & $\sim\!0.5$ & $9.38$\\
    N6  & $5501010104$ & $59816$ & $\sim\!4.0$ & $9.08$\\
    N7  & $5501010105$ & $59817$ & $\sim\!2.4$ & $9.32$\\
    N8  & $5501010106$ & $59818$ & $\sim\!2.5$ & $9.06$\\
    N9  & $5501010107$ & $59819$ & $\sim\!2.4$ & $9.16$\\
    N10 & $5501010108$ & $59820$ & $\sim\!3.0$ & $8.97$\\
    N11 & $5501010109$ & $59821$ & $\sim\!0.9$ & $9.45$\\
    N12 & $5501010110$ & $59822$ & $\sim\!3.9$ & $9.60$\\
    N13 & $5501010111$ & $59823$ & $\sim\!2.6$ & $9.18$\\
    \hline
    &\multicolumn{3}{c}{IXPE}&$1-10$ keV\\
    \hline
     X1 & $01250401$  & $59814$ & $\sim\!460$ & $4.03$\\ 
    \hline
    \end{tabular}}
    \end{table}

    \begin{figure}
    \centering
	    \includegraphics[trim={0 13mm 5mm 0}, clip,width=\columnwidth]{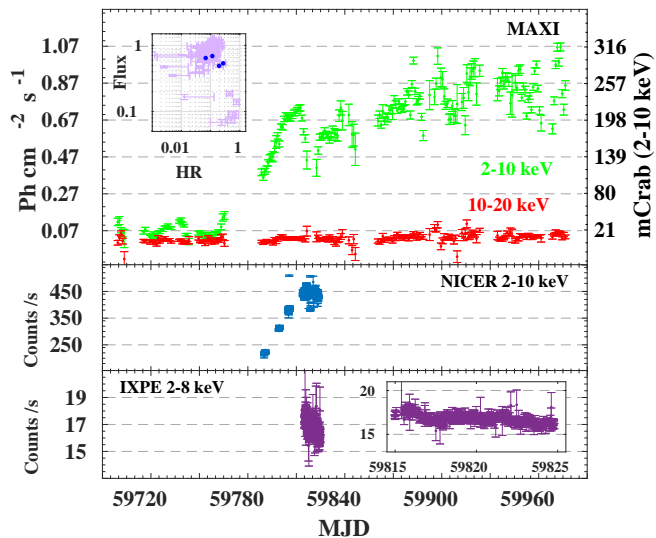}
        \caption{Top to bottom: 4U $1630-47$ light curves 
        obtained from {\it MAXI} (red and green),
        {\it NICER} (blue) and all three DUs of 
        {\it IXPE} (purple). The inset in the top panel
        shows the HID from {\it MAXI} with blue dots
        corresponding to {\it IXPE} observation duration.}
	\label{fig: all_lightcurve}	
	\end{figure}

    \subsection{IXPE}
    {\it IXPE} is an imaging polarimeter \citep{2022JATIS...8b6002W} 
    consisting of three detector units (DUs) which are 
    polarization sensitive in the $2-8$~keV energy range. 
    It observed 4U $1630-47$ from $23$ Aug, $2022$ 
    (MJD $59814$) to $02$ Sep, $2022$ (MJD $59824$) for
    $\sim\!460$ ksec (\autoref{tab: obs_table}).
    The Level-2 data of the observation are reduced and analysed 
    with {\tt IXPEOBSSIMv30.0.0} \citep{2022SoftX..1901194B}
    following the procedure of \citet{2023MNRAS.519.3681F}. 
    Further, {\tt XPSELECT} task is used to  extract the source 
    and background event lists. The source is considered within a 
    circular region of $60"$ and the background region as an annular region with 
    inner \& outer radii of $180"$ \& $240"$, respectively. 
    Subsequently, the {\tt XPBIN} task is
    used to generate the polarization cubes using the {\tt PCUBE} algorithm. 
    The Stokes I, Q and U spectra of source and background are generated
    with the {\tt PHA1}, {\tt PHA1Q} and {\tt PHA1U} algorithms.
    The light curves in $2-8$ keV energy band for the different 
    DUs are generated with the {\tt XSELECT} task of {\tt HEASOFT v6.31.1} 
    and added using the {\tt lcmath} task to get a 
    combined light curve (see bottom panel of \autoref{fig: all_lightcurve}).

    \subsection{NICER}
    {\it NICER} observed the source from $30$ Jul, $2022$ (MJD $59790$) 
    to $01$ Sep, $2022$ (MJD $59823$) which is well overlaps with 
    {\it IXPE} observations (see \autoref{tab: obs_table} and 
    \autoref{fig: all_lightcurve}). 
    Thirteen observations during this interval 
    are analyzed covering the rising phase of the outburst as well as 
    the peak of the same. The {\tt NICERDASv10} software distributed 
    with {\tt HEASOFT v6.31.1} is used along with the latest 
    {\tt CALDB} to reduce the data from the observations. The {\tt nicerl2} 
    task is used to perform standard calibration and screening to 
    generate cleaned event lists. The source and background spectra 
    along with the responses are generated in the $0.5-12$ keV energy 
    band using the {\tt nicerl3-spect} task. The spectra 
    are rebinned to have a minimum of $25$ counts per energy bin for
    spectral modelling. Further, the light curves for each observation 
    (\autoref{fig: all_lightcurve}) in the $2-10$ keV energy range are
    generated using the {\tt XSELECT} task. 

    \section{Modelling and Results}
    \label{sec: modelling and resuts}
    
    \subsection{Light Curve and Outburst profile }
	\label{sec: lightcurve}
    We consider one day averaged {\it MAXI} light curves 
    (see top panel of \autoref{fig: all_lightcurve}) 
    to get a comprehensive picture of the outburst 
    of 4U $1630-47$. The source, after $\sim\!150$ 
    days of inactivity goes into a new 
    outburst in the last few days of July $2022$. The {\it MAXI}
    count rate, in energy range of $2-10$ keV, increases from
    its quiescent value of $\sim\!0.07$ photons cm$^{-2}$ s$^{-1}$ 
    to $\sim\!10$ times on $21$ August, $2022$ (MJD $59812$). 
    At the time of writing this letter, the source is still is 
    in the outburst phase.  
    Moreover, the hardness intensity diagram (HID) generated with
    {\it MAXI} flux ($2-20$ keV) and the hardness ratio
    ($10-20$ keV/$2-10$ keV) for the entire outburst 
    duration represents an incomplete profile (see \autoref{fig: all_lightcurve})
    when compared with those of previous outbursts in $2016$ and $2018$ 
    \citep{2020MNRAS.497.1197B}. It suggests that the source 
    may remain in this phase for few more days before it 
    declines to quiescent low hard state (LHS).
    
    \subsection{Spectral modelling}
	\label{sec: diskbb fit}
	
\begin{figure}
	   \centering
		\includegraphics[trim={0 1mm 0 0}, clip,width=\columnwidth]{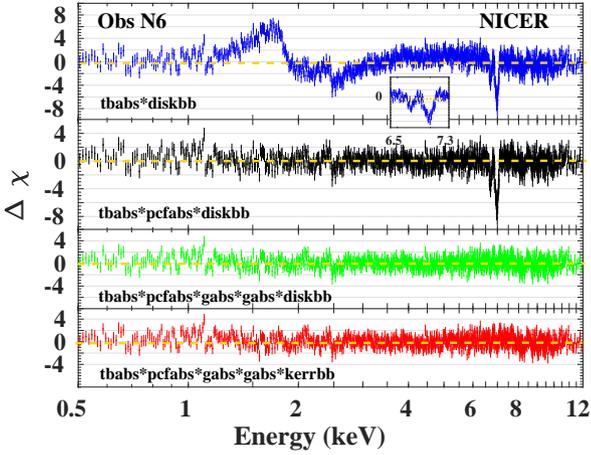}
	      \caption{The residuals (in units of $\sigma$) for several models applied to the 
        {\it NICER} spectra of 4U $1630−47$ obtained from Obs. N6. The corresponding 
        model is mentioned in each panel. The inset of top panel shows two absorption lines 
        present at $6.69$ keV and $6.97$ keV.}
	    \label{fig: nicer_spectra}
	\end{figure}

	The variability observed in {\it MAXI} light curve of 4U $1630-47$ 
    motivates us to investigate spectral nature and 
    its evolution during the outburst before we analyse the 
    polarization measurement from {\it IXPE}. Therefore, we model
    {\it NICER} spectra, in $0.5-12$ keV, of all the observations 
    (see \autoref{tab: obs_table}) starting from
    rising phase to peak of the outburst. 
    We began with Obs. N1 and fit a phenomenological model: 
    {\tt tbabs*diskbb}. The residuals from the fit do not
    suggest any requirement of high energy power-law but 
    an absorption below $4$ keV is noticed 
    (similar to top panel of \autoref{fig: nicer_spectra}). 
    Hence, a partial 
    covering component {\tt pcfabs} is multiplied in the model.
    This significantly improves the fit and results into 
    a $\chi_{red}^{2}= 1.10$. The instrument originated 
    characteristic Gold emission lines from X-ray optics of 
    {\it NICER} are corrected with inclusion of {\tt gauss} 
    components at $1.8$ and $2.2$ keV. 
    Further, two additional line absorption features are clearly visible 
    in the residual between $6$ to $7$ keV 
    ( similar to inset of \autoref{fig: nicer_spectra}) 
    and we model them with {\tt gabs} components. 
    These disk-wind originated absorption lines
    are reported previously and attributed to rest 
    frame energies of Fe XXV and Fe XXVI lines at $6.697$ keV 
    and $6.966$ keV, respectively (\citet{2014ApJ...784L...2K, 2018ApJ...867...86P}).
    Finally, the overall fit results $\chi_{red}^{2}$ = $1.02$ 
    and a disk temperature of kT$_{in}$ = $1.12\pm0.01$ keV
    with absorption lines at energies $6.69\pm0.01$ keV and 
    $6.97\pm0.01$ keV. The hydrogen column density N$_{H}$ 
    is estimated to be $(5.86\pm0.20)\times10^{22}$ atoms cm$^{−2}$. 
    The estimation of N$_{H}$ is slightly lower than that of 
    \citet{2018ApJ...867...86P, 2020MNRAS.497.1197B}.
    This, we attribute to the presence of {\tt pcfabs} in our model 
    which also has N$_{H}$ from covering column on top of 
    interstellar medium column of hydrogen atoms.
    Further, similar fit procedure is applied to rest of the 
    {\it NICER} observations from Obs. N2 to Obs. N13 with our 
    final phenomenological non-relativistic model
    {\tt tbabs*pcfabs*gabs*gabs*diskbb}. The residuals resulting at 
    various steps of fit procedure and from models are provided in \autoref{fig: nicer_spectra}
    from Obs. N6 for illustration.
    The spectral fits indicate a disk dominated state of the source
    and a very stable disk temperature of kT$_{in}$=$1.44\pm0.01$ keV 
    is observed during peak of the outburst phase. 

    \subsection{Spin estimation with Continuum Fitting (CF) method }
    \label{sec: kerrbb fit}
    
    The spectral modelling shows a disk dominated soft state 
    of the source during outburst. In such state, the energy spectra
    are suitable for spin estimation with CF method. Hence, we make use of
    {\it NICER } spectra, observed later than the peak of the 
    outburst (Obs. N4 to Obs. N13), to estimate the spin parameter 
    of BH. We follow \citet{2021MNRAS.507.2602K} to apply CF 
    method and model the spectra with the relativistic model: 
    {\tt tbabs*pcfabs*gabs*gabs*kerrbb}. The model results 
    into reasonably acceptable fits. The residuals, as previously,
    rule out any signature of power-law component in high energy 
    till $12$ keV (see bottom panel of \autoref{fig: nicer_spectra}). 
 
    We switch off limb-darkening and apply zero torque condition 
    at the inner boundary of the disk in {\tt kerrbb} component. 
    Moreover, the component is provided with the best estimations 
    of distance to the source (D), inclination of the binary plane ($i$)
    and mass of the BH (M$_{\rm{BH}}$) in the binary system  
    based on the work of \citet{2018ApJ...867...86P} 
    and \citet{2020MNRAS.497.1197B}. 
    Therefore, we fix D = $10$ kpc, $i = 65^{\circ}$, 
    M$_{\rm{BH}} = 10$ M$_{\odot}$. 
    The spectral hardening factor ($f$) is fixed to a fiducial 
    value of $1.55$ \citep{1995ApJ...445..780S,2018ApJ...867...86P}.
    The other two parameters of {\tt kerrbb} namely, 
    accretion rate ($\dot{\rm{M}}$) and spin parameter 
    (a$_{\ast}$) are allowed to vary freely.
    For Obs. N4, the overall fit estimates 
    N$_{H}=6.38^{+0.19}_{-0.17}\times10^{22}$ atoms cm$^{2}$, 
    a$_{\ast}=0.933^{+0.002}_{-0.003}\,(1\sigma)$ and 
    $\dot{\rm{M}}=1.62^{+0.04}_{-0.03}\times10^{18}$ g s$^{-1}$
    with $\chi^{2}_{red}=0.99~(898.05/908)$.
    Further, similar fit procedure is applied to rest 
    of the observations to estimate the spin parameter.
    The residuals of best fit model are shown in 
    \autoref{fig: nicer_spectra} from Obs. N6
    for illustration.
    We notice that the relativistic model gives better 
    fits (considering values $\chi^{2}_{red}$) than the non-relativistic model 
    and also results into consistent values of spin parameter
    from all the selected observations during the outburst of 
    4U $1630-47$ along with model generated uncertainties.  
    A detailed analysis with {\it NICER } data may be carried 
    out for estimation and error analysis on spin parameter
    \citep[see][]{2021MNRAS.507.2602K} which is out of scope of this work.
   
    \subsection{Polarimetric properties}
    \label{sec: polarimetric }
    
    \begin{figure}
    \centering
		\includegraphics[trim={0 1mm 0 0 }, clip, width =\columnwidth]{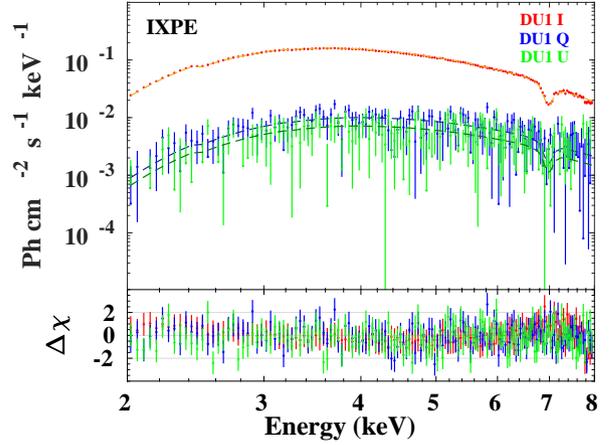}
	    \caption{{\it IXPE} energy spectra (unfolded) of 4U $1630-47$ during the outburst.
        Only DU1 Stokes I (red), Q (blue) and U (green) are shown along with fitted model
        (dashed line). Absorption feature at $6.9$ keV is also present.
        The bottom panel shows the residuals in units of $\sigma$.}
	    \label{fig: ixpe_spectra}
	\end{figure}
 
\begin{figure}
    \centering
        \includegraphics[trim={0 3mm 0 0},clip,width=0.6\columnwidth]{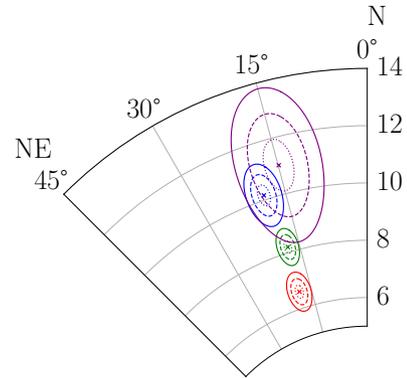}
        \caption{The confidence contours ($1\sigma$ [dotted], $2\sigma$ [dashed] and 
        $3\sigma$ [solid]) of PA and PD obtained with IXPE 
        in $2-3.5$ (red), $3.5-5$ (green), $5-6.5$ (blue) and $6.5-8$ keV (purple) energy bins. 
        The grid shows PA ($^{\circ}$) on radial lines and PD (\%) as concentric rings. 
        } 
        \label{fig: contour_plot}
     \end{figure}

    {\it IXPE} observed $\sim\!460$ ksec long portion of the outburst 
    of 4U $1630-47$. These observations spanned over $\sim\!10$ days
    (see \autoref{tab: obs_table}). 
    Simultaneous observations by {\it NICER} reveal a stable spectral 
    nature of system (see $\S$\ref{sec: diskbb fit} \&
    $\S$\ref{sec: kerrbb fit}). Hence, {\it IXPE} observations are very
    suitable to study polarimetric properties for the first time 
    from this source, using the methodology of \citet{2023MNRAS.519.3681F,
    2023arXiv230113394C, 2023arXiv230203396J}. Firstly, 
    the model-independent {\tt PCUBE} algorithm 
    \citep{2015APh....68...45K} is used to determine 
    the normalized Stokes parameters (Q/I \& U/I), polarization 
    angle (PA) and polarization degree (PD) in the 
    $2-3.5$ keV, $3.5-5$ keV, $5-6.5$ keV, $6.5-8$ keV 
    and in the total $2-8$ keV energy ranges, during the Obs. X1.
    The normalized stokes parameter are provided in 
    \autoref{tab: ixpe_stokes_table}. 
    PA and PD in four energy bands are given in 
    \autoref{tab: pcube_xspec_table} and corresponding 
    plots with error contours are presented in 
    \autoref{fig: contour_plot}. We note that the source 
    exhibits significant polarization, 
    with PD = $8.33\pm0.17$\% in $2-8$ keV energy band with all 
    the event data from three DUs combined. The PA in the 
    same energy band is found to be $17.78^{\circ}\pm0.60^{\circ}$.
    Moreover, we observe that PD has energy dependence whereas
    PA is roughly same over the energy range. 
    
    Subsequently, we also perform model-dependent analysis by 
    fitting {\it IXPE} Stokes I, Q and U spectra in {\tt XSPEC}. 
    We borrow the relativistic model as described in 
    $\S$\ref{sec: kerrbb fit} and insert the multiplicative component 
    {\tt polpow} which has PA and PD as function of energy. 
    The component defines PD(E) = A$_{\rm norm}\times$ E$^{\rm −A_{\rm index}}$ 
    and PA(E) = psi$_{\rm norm}\times$E$^{\rm −psi_{\rm index}}$. The
    {\tt polpow} is chosen over {\tt polcont} component based on
    the results of {\tt PCUBE} algorithm. 
    Hence, the model {\tt polpow*tbabs*gabs*gabs*pcfabs*kerrbb}
    is applied to fit three Stokes spectra from all the DUs 
    of {\it IXPE} at once in {\tt XSPEC} with the model. 
    Only parameters of {\tt polpow} component and 
    a$_{\ast}$, $\dot{\rm{M}}$ of {\tt kerrbb} are 
    kept free for the fit. All other 
    parameters are frozen to average values obtained 
    from {\it NICER} spectral fits of Obs. N4 to Obs. N13.
    The fit results ${\rm psi_{\rm index}}$, as nearly zero
    and hence, it is fixed to zero which does not affect the
    overall fit. Finally, the fit procedure results into 
    an acceptable $\chi_{red}^{2}=0.85$ for 
    combined nine spectra (I, Q \& U for 3 DUs). The best fit 
    values are provided in the \autoref{tab: fit_table}.
    \begin{table}
    \centering
    \caption{The best fit spectral parameters from {\it IXPE} 
    observations of 4U1630-47 with relativistic model.
     From left to right are,
			 (1) model components; 
		      (2) parameters in components;  
			     (3) best fit values for Obs. X1. The parameters 
            that are fixed  during the fits are denoted with $^{\rm fixed}$.}
	       \label{tab: fit_table}
    
        \begin{tabular}{lcc} 

		    \hline
			\hline
			Components 		&  \multicolumn{2}{c}{Parameter} \\
			\hline
			{\it polpow}    & A$_{\rm norm}$		    &$2.77\pm0.32$\\
                            & A$_{\rm index}$		    &$-0.76\pm0.07$\\
                            & psi$_{\rm norm}$		    &$17.72\pm0.53$\\
                            & psi$_{\rm index}$		    &$0^{\rm fixed}$\\
            {\it tbabs} & N$_{\rm H}$		    		&$5.84^{\rm fixed}$\\
			{\it pcfabs} & N$_{\rm H}$				    &$4.78^{\rm fixed}$\\
						   & CvrFract(\%)            &$86.94^{\rm fixed}$\\
			{\it kerrbb} & a$_{*}$					&$0.920\pm0.001$\\
			    		& $\dot{\rm{M}}$			&$1.50\pm0.01$\\
			    								
			                     
			\hline                     
		\end{tabular}
    \end{table}
    The overall fit of the spectra is shown in 
    \autoref{fig: ixpe_spectra}. Furthermore, the model-dependent 
    polarization results are compared with those 
    obtained in model-independent method by integrating PA(E) and PD(E) 
    in the $2-8$ keV energy band which results PA$\sim17.72^{\circ}$ and PD$\sim9\%$.
    We notice PA and PD from two different approaches
    of {\tt PCUBE} and {\tt XSPEC} are in good agreement. 
    We outline that the spin parameters resulting from 
    {\it NICER} spectral fits and {\it IXPE} polarimetric
    fits are also in good agreement.
    
  
    \begin{table}
    \centering
    \caption{Normalized Stokes parameters (with $1\sigma$ errors), 
    obtained from {\it IXPE} observation of 4U $1630-47$ computed 
    with the {\tt PCUBE} algorithm in four different energy bins.}
	   \label{tab: ixpe_stokes_table}
      \scalebox{0.88}{
	   \begin{tabular}{ccccc}
        \hline
        \hline
		& DU1           & DU2           & DU3            & All DUs                        \\
        \hline
		\multicolumn{5}{c}{$2-8$ keV}                                                   \\
        \hline
		Q/I (\%) & 6.74 $\pm$ 0.29 & 6.47 $\pm$ 0.30 & 7.14 $\pm$ 0.30  & 6.78 $\pm$ 0.17 \\
		U/I (\%) & 4.23 $\pm$ 0.29 & 4.96 $\pm$ 0.30 & 5.38 $\pm$ 0.30  & 4.85 $\pm$ 0.17 \\
        \hline
		\multicolumn{5}{c}{$2-3.5$ keV}                                                   \\
        \hline
		Q/I (\%) & 5.63 $\pm$ 0.40 & 4.46 $\pm$ 0.41 & 5.81 $\pm$ 0.42  & 5.30 $\pm$0.24   \\
		U/I (\%) & 3.45 $\pm$ 0.40 & 4.27 $\pm$ 0.41 & 4.04 $\pm$ 0.42  & 3.91 $\pm$ 0.24  \\
        \hline
		\multicolumn{5}{c}{$3.5-5$ keV}                                                   \\
        \hline
		Q/I (\%) & 6.29 $\pm$ 0.38 & 6.72 $\pm$ 0.39 & 7.12 $\pm$ 0.39  & 6.70 $\pm$ 0.22  \\
		U/I (\%) & 3.89 $\pm$ 0.38 & 4.71 $\pm$ 0.39 & 5.61 $\pm$ 0.39  & 4.71 $\pm$ 0.22  \\
        \hline
		\multicolumn{5}{c}{$5-6.5$ keV}                                                 \\
        \hline
		Q/I (\%) & 8.23 $\pm$ 0.64 & 8.38 $\pm$ 0.66 & 7.49 $\pm$ 0.66  & 8.04 $\pm$ 0.38  \\
		U/I (\%) & 5.43 $\pm$ 0.64 & 6.75 $\pm$ 0.66 & 6.52 $\pm$ 0.66  & 6.22 $\pm$ 0.38  \\
        \hline
		\multicolumn{5}{c}{$6.5-8$ keV}                                                   \\
        \hline
		Q/I (\%) & 9.33 $\pm$ 1.55 & 8.09 $\pm$ 1.64 & 11.30 $\pm$ 1.61 & 9.57 $\pm$ 0.92  \\
		U/I (\%) & 5.74 $\pm$ 1.55 & 4.18 $\pm$ 1.64 & 6.49 $\pm$ 1.61  & 5.48 $\pm$ 0.92  \\
        \hline
	   \end{tabular}}
    \end{table}

    \begin{table}
    \centering
    \caption{PD and PA (with $1\sigma$ errors), obtained from 
    {\it IXPE} observation of 4U $1630-47$ computed with 
    the {\tt PCUBE} algorithm in different energy bins }
	    \label{tab: pcube_xspec_table}
        \scalebox{0.79}{
	       \begin{tabular}{lccccl}
        \hline
        \hline
		& $2-8$ keV & $2-3.5$ keV  & $3.5-5$ keV  & $5-6.5$ keV & $6.5-8$ keV  \\
        \hline
        \multicolumn{6}{c}{\tt PCUBE}\\
        \hline
        PD (\%) & 8.33 $\pm$ 0.17 & 6.58 $\pm$ 0.24 & 8.19 $\pm$ 0.22 & 10.17 $\pm$ 0.38 & 11.02 $\pm$ 0.92\\ 
        PA ($^{\circ}$)& 17.78 $\pm$ 0.60 & 18.21 $\pm$ 1.02 & 17.56 $\pm$ 0.78 & 18.87 $\pm$ 1.06 & 14.90 $\pm$ 2.40\\
        \hline
        \end{tabular}}
    \end{table}
   
	\section{Discussion and Conclusion}
	\label{sec: discussion}

    In this letter, we report the significant detection of X-ray 
    polarization from a recurrent transient BH-XRB 4U $1630-47$
    by {\it IXPE} during its outburst in $2022$. The polarization 
    properties are consolidated with simultaneous spectral 
    studies by {\it NICER} observations.

    The most intriguing finding from our investigation is 
    a substantial degree of polarization (PD) = $8.33\pm0.17$\% 
    ($>48\,\sigma$ statistical confidence) in disk dominated 
    outburst phase of the source in $2-8$ keV energy range. 
    The energy resolved PD reaches as high as $11.02\pm0.92\%$
    ($>11\,\sigma$ statistical confidence) in $6.5-8$ keV. 
    These measurements are significantly higher when 
    compared to those predicted from existing models 
    \citep{2008MNRAS.391...32D, 2009ApJ...701.1175S, 2020MNRAS.493.4960T}.
    Although the increasing nature of PD within $2-8$ keV 
    energy range is similar as depicted by the models.
    This indicates that additional physical processes and effects
    are in play on top of electron scattering on the surface of disk 
    along with reflection of return radiation from the near side of 
    the disk which can further enhance the PD. 
        
    Recently, \citet{2022Sci...378..650K} reported  
    polarization properties of Cygnus X-1 in LHS,
    and found that the observed PD is approximately twice as high
    as expected. These polarimetric results favour 
    the presence of the coronal plasma in wedge shaped 
    geometry with in the system. In the case of 
    4U $1630-47$, such strong Compton 
    scattering medium like corona may not be present as the 
    {\it NICER} spectra during the outburst show no signature 
    of high energy tail. But, a weak corona may be present 
    as observed in previous outbursts 
    \citep{2018ApJ...867...86P, 2020MNRAS.497.1197B}.
    A weak corona may contribute to some extent in the
    net observed PD.

    Another plausible contribution in PD could be from scattering 
    of the photons in an obscuring medium generated by disk-wind 
    between inner regions of the disk and an observer. The strong evidence 
    of disk-wind comes from the {\it NICER} spectral modelling. 
    The spectra during the outburst exhibit presence of partial 
    covering over the disk radiation and absorption line features
    of Fe XXV and Fe XXVI at rest frame energies 
    $6.697$ keV and $6.966$ keV respectively (see \autoref{fig: nicer_spectra}).
    These absorption features are also present in {\it IXPE}-I spectrum 
    but the two lines are merged, resulting in a broad absorption dip 
    (see \autoref{fig: ixpe_spectra}). Due to poor statistics, the presence of 
    the absorption feature is not clear in IXPE-Q and U spectra. Although
    the difference in intensities of the absorption lines, if they were present,
    in Q and U spectra would have provided an important piece of evidence 
    about the origin of polarized emission from disk-wind. Recently, similar 
    absorption features associated with disk-wind are also observed 
    in outburst of 4U $1543-47$ \citep{2023MNRAS.tmp..141P}.
    
    Disk-wind in soft states
    may have a column density up to N$_{H}\sim\!10^{25}$ cm$^{-2}$ in 
    the equatorial direction. A rough qualitative estimates are 
    provided and discussed by \citet{2021A&A...646A.154R}. 
    The wind may be fully ionized but Compton thick in 
    the equatorial direction. The effective column density and 
    therefore PD may vary with inclination angle ($i$) with 
    respect to a distant observer. Hence, the polarized 
    signature of 4U $1630-47$ may be affected 
    by disk-wind and its geometry.

    We also estimate the associated electric field angle 
    as PA = $17.78^{\circ}\pm0.60^{\circ}$ which remains
    mostly same within the $2-8$ keV energy range of {\it IXPE}.
    The paucity of radio observations and measurements 
    of jet angle from the source prevented us to relate it
    with geometry of the system. Although we note that the PA 
    is comparable to $90^{\circ}$ minus the inclination
    of the binary plane. In case of Cygnus X-1, PA
    is found to align with jet angle, which is assumed
    to be parallel to BH spin axis or normal to disk plane
    \citep{2022Sci...378..650K}.

    Furthermore, PD and PA are estimated with spectro-polarimetric
    modelling in {\tt XSPEC} as $\sim9\%$ and 
    $\sim17.72^{\circ}$, respectively.
    These model-dependent estimations of PD and PA are in good 
    agreement with those computed from model-independent {\it PCUBE} 
    algorithm. The values lie within $3\sigma$ uncertainties. 
    
    The spectro-polarimetric fit of {\it IXPE} data with 
    relativistic model also constrains the BH spin 
    parameter: a$_{\ast}=0.920\pm0.001\,(1\sigma)$ and accretion rate:
    $\dot{\rm{M}}=(1.50\pm0.01)\times10^{18}$ g s$^{-1}$.
    The $\dot{\rm{M}}$ is shown to be related to the mass 
    loss rate through disk-wind outflow in 4U $1630-47$ and other
    BH-XRBs \citep{2011ApJ...737...69N, 2012MNRAS.422L..11P}.
    
    We compare the value of a$_{\ast}$ with that obtained
    by applying CF method on {\it NICER} spectra during peak of
    the outburst which resulted a$_{\ast}=0.933^{+0.002}_{-0.003}\,(1\sigma)$.
    The spin parameter estimated with two different approaches
    are in good agreement. The estimations are also in line
    with findings of \citet{2018ApJ...867...86P}, whereas
    \citet{2014ApJ...784L...2K} indicated presence of a
    maximally rotating BH in the system. Recent measurement
    by {\it Insight-HXMT} suggests a slow rotating BH with
    a$_{\ast}=0.817\pm0.014$ ($90\%$ statistical error) 
    \citet{2022MNRAS.512.2082L}.
    Here, we highlight the fact that estimation 
    of spin parameter is highly dependent on $i$, D and M. 

    To summarize, we report a significant detection of polarized emission from
    4U $1630-47$ during its outburst in $2022$. The observed high degree of polarization 
    may be due to combined effects from the accretion disk, disk-wind and a weak corona
    in the system. The spin of the black hole is also estimated with spectro-polarimetric
    data of {\it IXPE}.
    
    \section*{Acknowledgements}
    We thank the anonymous reviewer for 
    careful reading of our manuscript to provide
    insightful comments and suggestions that significantly 
    improved the manuscript in terms of both science as well as
    writing. Authors thank GH, SAG; DD, PDMSA and Director, URSC for 
	encouragement and continuous support to carry out this 
    research. {\it IXPE}, {\it NICER} and {\it MAXI} teams
    are also thanked for providing data products and software tools
    for data analysis.
        
    \section*{Data Availability}
    Data used for this work are available at 
	{\it HEASARC} website 
    (\url{https://heasarc.gsfc.nasa.gov/docs/archive.html})
	and {\it MAXI} website
	(\url{http://maxi.riken.jp/top/index.html}).
	

    \bibliographystyle{mnras}
    \bibliography{references} 

    \bsp	
    \label{lastpage}
    \end{document}